# Numerical study of the thermoelectric power factor in ultra-thin Si nanowires

Neophytos Neophytou and Hans Kosina

Institute for Microelectronics, TU Wien, Gußhausstraße 27-29/E360, A-1040 Wien, Austria

e-mail: {neophytou|kosina}@iue.tuwien.ac.at

## Abstract

Low dimensional structures have demonstrated improved thermoelectric (TE) performance because of a drastic reduction in their thermal conductivity, $\kappa_l$. This has been observed for a variety of materials, even for traditionally poor thermoelectrics such as silicon. Other than the reduction in $\kappa_l$, further improvements in the TE figure of merit $ZT$ could potentially originate from the thermoelectric power factor. In this work, we couple the ballistic (Landauer) and diffusive linearized Boltzmann electron transport theory to the atomistic $sp^3d^5s^*$-spin-orbit-coupled tight-binding (TB) electronic structure model. We calculate the room temperature electrical conductivity, Seebeck coefficient, and power factor of narrow 1D Si nanowires (NWs). We describe the numerical formulation of coupling TB to those transport formalisms, the approximations involved, and explain the differences in the conclusions obtained from each model. We investigate the effects of cross section size, transport orientation and confinement orientation, and the influence of the different scattering mechanisms. We show that such methodology can provide robust results for structures including thousands of atoms in the simulation domain and extending to length scales beyond 10nm, and point towards insightful design directions using the length scale and geometry as a design degree of freedom. We find that the effect of low dimensionality on the thermoelectric power factor of Si NWs can be observed at diameters below ~7nm, and that quantum confinement and different transport orientations offer the possibility for power factor optimization.

# I. Introduction

The ability of a material to convert heat into electricity is measured by the dimensionless figure of merit $ZT=\sigma S^2 T/(\kappa_e+\kappa_l)$, where $\sigma$ is the electrical conductivity, $S$ is the Seebeck coefficient, and $\kappa_e$ and $\kappa_l$ are the electronic and lattice part of the thermal conductivity, respectively. The interrelation between $\sigma$, $S$, and $\kappa_e$ in bulk materials keeps $ZT$ low [1]. Some of the best thermoelectric materials are compounds of Bi, Te, Pb, Sb, Ag, and exhibit $ZT \sim 1$ [1, 2]. Recently, however, using low-dimensional structures, it was demonstrated that $ZT$ could be greatly increased compared to their bulk counterparts, setting the stage for highly efficient TE energy conversion.

It was initially suggested that thermoelectric efficiency could be improved at the nanoscale because of two reasons: i) Low-dimensionality and quantum size effects could improve the Seebeck coefficient [3], and ii) Small feature sizes enhance phonon scattering on nanoscale interfaces and reduce thermal conductivity [4]. Indeed, large improvements of $ZT$ in low-dimensional structures such as 0D quantum dots, 1D nanowires (NWs), 2D superlattices and bulk nanocomposites have recently been achieved [4, 5, 6, 7, 8, 9, 10, 11, 12, 13, 14]. This was even achieved for common materials, and importantly Si based systems such as Si, SiGe, and SiC [15, 12, 13, 14]. Silicon, the most common semiconductor with the most advanced industrial processes, is a poor TE material with $ZT_{bulk}\sim0.01$. Si NWs, on the other hand, have demonstrated $ZT\sim1$, a 100X increase [12, 13, 14, 15], and they are now considered as emerging candidates for high efficiency and large volume production TE applications [16].

Most of the benefit to the measured $ZT$ values of NWs originates from a dramatic reduction in the lattice thermal conductivity $\kappa_l$ [15, 17, 18, 19, 20]. It has very recently become evident, however, that benefits from $\kappa_l$ reduction are reaching their limits, and further increases of $ZT$ can only be achieved through improvements in the power factor $\sigma S^2$ [16, 21]. By nanostructuring, the electronic structure could be engineered to tune the Seebeck coefficient [3, 7, 22] and the electrical conductivity [23] independently, which could maximize $\sigma S^2$. For example, Hicks and Dresselhaus suggested that the sharp features in the low-dimensional density of states function *DOS(E)* can improve the Seebeck coefficient [3, 7]. Mahan and Sofo have further shown that thermoelectric energy conversion through a single energy level (0D channel) can reach the Carnot efficiency when $\kappa_l$ is zero [24]. Because of the strong interconnection between $\sigma$ and $S$, and their dependence on the geometrical features, involved simulation capabilities that account for the atomistic nature over large length scales are necessary in order to guide the design of such devices.

In this work the atomistic $sp^3d^5s^*$-spin-orbit-coupled ($sp^3d^5s^*$-SO) tight-binding model [25, 26, 27, 28, 29] is used to calculate the electronic structure of thin silicon NWs. Two transport formalisms are employed to calculate the thermoelectric coefficients $\sigma$, $S$, and the power factor $\sigma S^2$: i) The Landauer formalism [30, 31, 32, 33, 34], and linearized Boltzmann theory [23, 24, 35]. We describe the numerical methodologies and the approximations used, and demonstrate why such methodology is appropriate and efficient for this purpose. We consider different NW diameters, different transport orientations ([100], [110], [111]), different cross section geometries and various relevant scattering mechanisms. Using experimental values for $\kappa_l$ in Si NWs, we estimate the *ZT* figure of merit. Our results explore effects of bandstructure features resulting from scaling the channel cross sections on the TE coefficients. Design optimization directions based on bandstructure engineering in low-dimensional channels are identified.

The paper is organized as follows: In section II we describe the Landauer approach which is used to investigate the effect of the geometrical features on the electronic structures and the thermoelectric coefficients of ultra-scaled Si NWs. In section III we describe the numerical approach to couple the TB model and Boltzmann transport theory, and the approximations used. In section IV we investigate the effects of NW cross section size, orientation, and scattering mechanisms on the thermoelectric coefficients. Finally, in V we conclude.

## II. Ballistic Landauer approach for TE coefficients

The NW bandstructure is calculated using the 20 orbital atomistic tight-binding sp$^3$d$^5$s*-SO model [25, 28], which is sufficiently accurate and inherently includes the effects of different transport and quantization orientations. We consider infinitely long, uniform, silicon NWs in the [100], [110] and [111] transport orientations as shown in Fig. 1, with different cross section shapes. We assume passivated surfaces. The passivation technique details are provided in Appendix 2 [36]. These geometrical features have an impact on the electronic structure and the transport properties. Figure 2 shows examples of n-type NW electronic structures. The lowest subbands are shifted to the same origin *E=0eV* for comparison purposes. For brevity, only half of the *k*-space is shown. Figure 2a and Fig. 2b show the dispersions of [100] NW with diameters of 3nm and 12nm, respectively. As the diameter is reduced, the number of subbands is reduced, and the relative shift between the Γ and off-Γ valleys also changes. The degeneracies ($\eta$) of the Γ and off- Γ valleys are $\eta = 4$ and $\eta = 1$, respectively. Figure 2c and Fig. 2d show the corresponding dispersions for the [111] NW. The shape of the dispersion is different for different orientations. The degeneracy of this valley is $\eta =6$. For [110] NWs, Fig. 2e shows the dispersion of the 3nm wide and 12nm tall rectangular NW (strong (1-10) surface quantization), whereas Fig. 2f the dispersion of the 12nm wide and 3nm tall NW (strong (001) surface quantization), which produce different electronic structures. The degeneracy of both, the Γ and off- Γ valleys is $\eta =2$. This dependence of the dispersions on geometry will result in different electronic and thermoelectric characteristics.

In this section, the ballistic Landauer formalism [30] is used to extract the TE coefficients. Although ballistic transport cannot be achieved in a realistic thermoelectric device, the results in this section indicate the upper performance limit, and it is a "fast" way to identify whether geometry could have an effect on TE properties through bandstructure engineering. The results from this method are compared to the results from the diffusive Boltzmann transport method in section III.

In the Landauer formalism the current is given by:

$$J = -\frac{q_0}{L}\sum_{k>0} v_k f_1 - \frac{q_0}{L}\sum_{k<0} v_k f_2 \tag{1a}$$

$$= -\frac{q_0}{L}\sum_{k>0} v_k (f_1 - f_2), \tag{1b}$$

where $v_k$ is the bandstructure velocity, and $f_1$, $f_2$ are the Fermi functions of the left and right contacts, respectively. Auxiliary functions $R^{(\alpha)}(f_1, f_2, T)$ can be defined as:

$$R^{(\alpha)} = \frac{-\dfrac{q_0}{L}\sum_{k>0} v_k (f_1 - f_2)(E_k - \mu_1)^\alpha}{(\mu_1 - \mu_2)}, \tag{2}$$

where $\mu_1$, $\mu_2$ are the contact Fermi levels, and $E_k$ is the subband dispersion relation. This formula is the same as the one described in references [31, 37], where for small driving fields $\Delta V$, the linearization $f_1 - f_2 = -q_0 \Delta V \dfrac{\partial f_1}{\partial E}$ is applied. Here, however, the computation is explicitly performed in $k$-space rather than energy-space. From these functions, the conductance $G$, the Seebeck coefficient $S$, and the electronic part of the thermal conductivity $\kappa_e$, can be derived as

$$G = R^{(0)}, \tag{3a}$$

$$S = \frac{1}{T}\frac{R^{(1)}}{R^{(0)}}, \tag{3b}$$

$$\kappa_e = \frac{1}{T}\left[R^{(2)} - \frac{\left[R^{(1)}\right]^2}{R^{(0)}}\right]. \tag{3c}$$

Using this approach, the power factor (defined as $\sigma S^2 = G/\text{Area}*S^2$) has been calculated. It is shown in Fig. 3 as a function of the one-dimensional carrier concentration, for cylindrical n-type NWs in the three transport orientations [100], [110] and [111] for two different diameters $D=3$nm and $D=12$nm. Comparing the magnitude of the power factor for $D=3$nm, the [111] NW with a 6-fold degenerate band has a higher power factor than the other NWs. The [100] NW, with a 4-fold degenerate Γ-valley follows, whereas the [110] NW with a 2-fold degenerate Γ-valley has the lowest power factor. Subbands with higher degeneracies, or subbands with edges very close in energy, improve the Seebeck coefficient which can be beneficial to the power factor. We show in section III, however, that once scattering is included in the calculation, the conductivity is degraded, which turns out to be a more dominant effect than the increase in Seebeck coefficient. For *D*=12nm in Fig. 3, the NW bandstructure becomes bulk-like, and any orientation effects that existed because of the bandstructure differences in lower diameters, are now smeared out. The interesting observation, however, is that under ballistic conditions, it seems that it is possible to improve the thermoelectric power factor by feature size scaling, in agreement with other theoretical ballistic transport studies, [31, 32, 33, 34]. The magnitude of these benefits, however, is only within a factor of two.

To also emphasize the effect of the different confinement orientations, Fig. 4a shows the power factor for n-type [100] NWs, as a function of the carrier concentration for different confinement conditions for a *square* NW. Starting from the 12nm x 12nm "bulk-like" NW, we examine two cases: (i) we reduce the size of one of the sides, i.e. the height *H* is scaled to *H*=3nm, while the width *W* is kept at *W*=12nm in decrements of 1nm (red lines). This represents the case of scaling from bulk towards "thin-body" devices. (ii) We scale both the width *W* and height *H* simultaneously down to *W*=*H*=3nm in decrements of 1nm (blue lines). In both cases, decreasing the feature size of either side increases the peak of the power factor. The increase is larger when both sides are scaled, noted (3,3). In this case, cross section scaling is beneficial for the power factor. Those benefits are ~50%, and appear for side sizes below ~7nm (for sizes above that the power factor saturates).

Figure 4b, shows the same features for the n-type [110] NWs. The sides are [1-10] in the width and [001] in the height directions. Two device families are shown: (i) Devices with constant width along [1-10] at *W*=3nm, while the height along [001] varies from *H*=3nm to *H*=12nm (thin and tall NWs – red lines). (ii) Devices with the reverse aspect ratio, for which *W* varies from *W*=3nm to 12nm, while *H* is fixed at *H*=3nm (wide and thin NWs – blue lines). The peaks of the power factors of the first device series (red lines) are higher than those of the second device series (blue lines). Interestingly, they are even higher than the peak of the fully scaled 3nm x 3nm NW, indicating that cross section scaling is not always beneficial, even for ballistic channels. The relative

performance in these channels, as in the case of the ones described in Fig. 3, originate from the higher Seebeck coefficient, which is a consequence of the larger number of subbands/degeneracies in the electronic structure of this nanowire near the conduction band edge. The 3nm x 12nm NW has a higher performance than the 12nm x 3nm because, as previously shown in Fig. 2e, the band edges of the $\Gamma$ and off-$\Gamma$ valleys are nearby in energy. For the 12nm x 3nm NW in Fig. 2f, the off-$\Gamma$ valleys are higher in energy and do not participate in transport.

Figure 4c and Fig. 4d show the figure of merit *ZT* values of the devices in Fig. 4a and Fig. 4b, respectively, using a single value of $\kappa_l$=2W/mK for the lattice part of the thermal conductivity, which was experimentally demonstrated for NWs [13, 15, 18]. *ZT* follows the shape and trends of the power factor. Interestingly, under ballistic assumptions, very high *ZT* values up to 4 can be achieved. We emphasize that such a low value for the thermal conductivity has experimentally only been achieved in rough or distorted NWs. We still use it, however, although for electrons we consider ballistic transport. Our intention here is to provide an idealized upper value for the *ZT* in Si NWs. As we will describe later on in section IV, such values cannot be obtained once surface roughness scattering is incorporated. On the other hand, other methods for achieving very low thermal conductivity values have been theoretically proposed, which do not rely on surface roughness. Markussen *et* al., has proposed that Si nanowires, having surfaces decorated with molecules could also significantly reduce thermal conductivity, for which case our results are more relevant [38].

Another possibility to further improve thermoelectric performance is by adjusting the band positioning through gating. The gate electric field, similar to transistor devices, could shift the bands and change the thermoelectric properties. Figure 5 demonstrates this effect. Figure 5a shows the electronic structure of the *D*=12nm [111] n-type nanowire under flat potential in the cross section, whereas Fig. 5b under high gate inversion conditions. The separation of the bands has changed, and this results in an improvement of the thermoelectric ballistic *ZT* value by ~40%, which is a significant improvement. Careful design of the subband placement is, therefore, needed for improved performance.

The nanostructure geometry enters the design through subband engineering. The tight-binding (TB) model is particularly suited for this, because the computational domain can be extended beyond 10nm, and the effect of length scale can be properly investigated.

## III. Linearized Boltzmann approach for TE coefficients

The ballistic Landauer approach emphasizes the effect of the Seebeck coefficient through subband positioning, whereas the conductivity of the channel is not affected by the otherwise enhanced scattering in ultra-narrow channels. In this section, we describe an approach to couple the TB model to linearized Boltzmann transport theory in order to investigate thermoelectric (TE) properties in 1D Si NWs in the diffusive transport regime. Several approximations are made in order to make the computation more robust, without affecting the essence of the conclusions. The entire procedure is described in detail in our previous works [23, 39]. Here, we only present the basic formalism, but we focus on the numerical and computational details of the method.

In Linearized Boltzmann theory, the TE coefficients are defined as:

$$\sigma = q_0^2 \int_{E_0}^{\infty} dE \left( -\frac{\partial f_0}{\partial E} \right) \Xi(E), \tag{4a}$$

$$S = \frac{q_0 k_B}{\sigma} \int_{E_0}^{\infty} dE \left( -\frac{\partial f_0}{\partial E} \right) \Xi(E) \left( \frac{E - E_F}{k_B T} \right), \tag{4b}$$

$$\kappa_0 = k_B^2 T \int_{E_0}^{\infty} dE \left( -\frac{\partial f_0}{\partial E} \right) \Xi(E) \left( \frac{E - E_F}{k_B T} \right)^2, \tag{4c}$$

$$\kappa_e = \kappa_0 - T\sigma S^2. \tag{4d}$$

The transport distribution function $\Xi(E)$ is defined as [24, 35]:

$$\Xi(E) = \frac{1}{A} \sum_{k_x, n} v_n^2(k_x) \tau_n(k_x) \delta(E - E_n(k_x))$$
$$= \frac{1}{A} \sum_n v_n^2(E) \tau_n(E) g_{1D}^n(E). \tag{5}$$

where $v_n(E) = \frac{1}{\hbar}\frac{\partial E_n}{\partial k_x}$ is the bandstructure velocity, $\tau_n(k_x)$ is the momentum relaxation time for a carrier with wave-number $k_x$ in subband $n$, and

$$g_{1D}^n(E_n) = \frac{1}{2\pi\hbar}\frac{1}{|v_n(E)|} \qquad (6)$$

is the density of states for 1D subbands (per spin). The transition rate $S_{n,m}(k_x, k_x')$ for a carrier in an initial state $k_x$ in subband $n$ to a final state $k_x'$ in subband $m$ is extracted from the atomistic dispersions using Fermi's Golden Rule [40]:

$$S_{n,m}(k_x, k_x') = \frac{2\pi}{\hbar}\left|H_{k_x',k_x}^{m,n}\right|^2 \delta\left(E_m(k_x') - E_n(k_x) - \Delta E\right). \qquad (7)$$

Usually, the momentum relaxation times are calculated by:

$$\frac{1}{\tau_n(k_x)} = \sum_{m,k_x'} S_{n,m}(k_x, k_x')\left(1 - \frac{|p_m(k_x')|}{|p_n(k_x)|}\cos\vartheta\right) \qquad (8)$$

where in 1D the angle $\vartheta$ can take only two values $\vartheta = 0$ and $\vartheta = \pi$ [40, 41].
In this work, we calculate the relaxation times by:

$$\frac{1}{\tau_n(k_x)} = \sum_{m,k_x'} S_{n,m}(k_x, k_x')\left(1 - \frac{v_m(k_x')}{v_n(k_x)}\right) \qquad (9)$$

Both are simplifications of the actual expression that involves an integral equation for $\tau_n$ [41, 42, 43, 44]:

$$\frac{1}{\tau_n(k_x)} = \sum_{m,k_x'} S_{n,m}(k_x, k_x')\left(1 - \frac{v_m(k_x')\tau_m(k_x')f_m(k_x')}{v_n(k_x)\tau_n(k_x)f_n(k_x)}\right) \qquad (10)$$

While self-consistent solutions of this equation may be found, this is computationally very expensive, especially for atomistic calculations. Therefore, it is common practice in the literature to simplify the problem [45, 46, 47, 48], and often sufficiently accurate results are obtained using the above approximations [42, 43]. For a parabolic dispersion, the use of Eq. 8 and Eq. 9 is equivalent. For a generalized dispersion, however, where the effective mass of the subbands is not well defined and the valleys appear in various places in the Brillouin zone, and the use of Eq. 9 is advantageous.

The matrix element between a carrier in an initial state $k_x$ in subband $n$ and a carrier in a final state $k_x^{'}$ in subband $m$ is defined as:

$$H^{m,n}_{k_x^{'},k_x} = \frac{1}{\Omega} \int_{-\infty}^{\infty} \int_R F_m(\vec{R})^* e^{-ik_x^{'}x} U_S(\vec{r}) F_n(\vec{R}) e^{ik_x x} d^2R dx, \qquad (11)$$

where the total wavefunction is decomposed into a plane wave $e^{ik_x x}$ in the x-direction, and a bound state $F_v(\vec{R})$ in the transverse, in-plane, with $\vec{R}$ being the in-plane vector. $U_S(\vec{r})$ is the scattering potential and $\Omega$ is the normalization volume. We note here that Eq. (11), and later on Eq. (14) and Eq. (23) involve integrals of the function $F_{m/n}$ over the NW in-plane $R$. However, $F_{m/n}$ is only sampled on the atomic sites. In the actual calculation the integrals over $R$ are converted to summations over the atomic sites. The procedure is described in detail in Appendix 1.

Elastic and inelastic scattering processes are taken into account. We consider bulk phonons and following the same rules when selecting the final states for scattering as in bulk Si. For n-type nanowires (NWs), the elastic processes due to elastic acoustic phonons, surface roughness (SRS), and impurity scattering are only treated as intra-valley processes, whereas inelastic processes due to inelastic phonons are only treated as inter-valley (IVS). An example of such transitions is shown in Fig. 6 for the $D$=3nm [110] NW. Although all valleys from the bulk Si electronic structure collapse from 3D to 1D k-space in our calculations, we carefully chose the final scattering states for each event by taking into account the degeneracies of the projected valleys for each orientation differently, as also indicated in Fig. 6. For inelastic transitions all six *f*- and *g*-type processes are included [40, 49]. For p-type NWs we consider ADP (acoustic deformation potential) and ODP (optical deformation potential) processes which can be intra-band and inter-band as well as intra-valley and inter-valley.

For the scattering rate calculation, we extend the usual approach for 3D and 2D thin-layer scattering commonly described in the literature [40], to 1D electronic structures. For phonon scattering, the relaxation rate of a carrier in a specific subband $n$ as a function of energy is given by [23, 39]:

$$\frac{1}{\tau_{ph}^n(E)} = \frac{\pi}{\hbar} \frac{\left(N_\omega + \frac{1}{2} \mp \frac{1}{2}\right)}{\rho \hbar \omega_{ph}} \\ \times \left(\frac{1}{L_x} \sum_{m,k_x'} \frac{|K_{\vec{q}}|^2}{A_{nm}^{k_x k_x'}} \delta_{k_x',k_x \pm q_x} \delta\left(E_m(k_x') - E_n(k_x) \pm \hbar \omega_{ph}\right)\left(1 - \frac{v_m(k_x')}{v_n(k_x)}\right)\right), \quad (12)$$

where $\hbar \omega_{ph}$ is the phonon energy, and we have used $\Omega = AL_x$. For optical deformation potential scattering (ODP for holes, IVS for electrons) it holds $|K_{\vec{q}}|^2 = D_O^2$, whereas for acoustic deformation potential scattering (ADP or IVS) it holds $|K_{\vec{q}}|^2 = q^2 D_{ADP}^2$, where $D_0$ and $D_{ADP}$ are the scattering deformation potential amplitudes. Specifically for elastic acoustic deformation potential scattering (ADP), after applying the equipartition approximation, the relaxation rate becomes:

$$\frac{1}{\tau_{ADP}^n(E)} = \frac{2\pi}{\hbar} \frac{D_{ADP}^2 k_B T}{\rho v_s^2} \left(\frac{1}{L_x} \sum_{m,k_x'} \frac{1}{A_{k_x k_x'}^{nm}} \delta_{k_x',k_x \pm q_x} \delta\left(E_m(k_x') - E_n(k_x)\right)\left(1 - \frac{v_m(k_x')}{v_n(k_x)}\right)\right), \quad (13)$$

where $v_s$ is the sound velocity in Si.

In the expressions above, the quantities in the right-hand-side are all *k*-resolved when computed from the electronic structure *E(k)*, whereas the scattering rate in the left-hand-side is a function of energy. The δ-function in Eq. (12) and (13) states energy conservation. Numerically, the *E(k)* relation needs to be discretized in energy. All states are sorted in energy. and at a particular energy, arrays with all relevant *k*-states from all subbands are constructed.

One of the computationally most demanding steps in terms of memory requirements is the calculation of $A_{k_x k_x'}^{nm}$, the wavefunction overlap between the final and initial states. The calculation of this quantity involves an integral of the form:

$$\int_R \rho_{k_x',k_x}^{m,n}(\vec{R})^2 d^2R, \quad (14b)$$

$$\rho_{k_x',k_x}^{m,n}(\vec{R}) = F_{m,k_x'}(\vec{R})^* F_{n,k_x}(\vec{R}). \tag{14b}$$

For the larger NWs, this calculation of the matrix elements imposes a huge computational burden. All wavefunctions of every *k*-state for every subband need to be stored because it is not known *a priori* for each initial state which are the corresponding final scattering *k*-states and at which subbands when calculating the electronic structure, as indicated in the scattering examples of Fig. 6. The *D*=12nm structures that could include 5500 atoms each described by 20 orbitals, and a typical *k*-space grid of 200 points and considering 100 subbands, require several tens of Gbytes for the storage of the wavefunctions alone. For computational efficiency, therefore, we use the following scheme: on each atom we add the probability density of the components of each multi-orbital wavefunction, and afterwards perform the final/initial state overlap multiplication. In such way, we approximate the form factor components of a lattice atom at a specific location $R_0$ by:

$$\begin{aligned} \left|\rho_{k_x',k_x}^{m,n}\right|^2 &= \sum_\alpha F_{n,k_x}^\alpha F_{m,k_x'}^{\alpha\,*} \sum_\beta F_{n,k_x}^\beta F_{m,k_x'}^{\beta\,*} \\ &\approx \sum_{\alpha,\beta} F_{n,k_x}^\alpha F_{n,k_x}^{\beta\,*} \delta_{\alpha,\beta} \sum_{\alpha,\beta} F_{m,k_x'}^\alpha F_{m,k_x'}^{\beta\,*} \delta_{\alpha,\beta} \\ &= \sum_\alpha \left|F_{n,k_x}^\alpha\right|^2 \sum_\beta \left|F_{m,k_x'}^\beta\right|^2 \equiv \left\langle F_{n,k_x} \right\rangle^2 \left\langle F_{m,k_x'} \right\rangle^2 \end{aligned} \tag{15}$$

where $\alpha, \beta$ run over the tight-binding orbitals of a specific atom. With this the overlaps are computed using the probability density of each state, as in a single orbital (i.e. effective mass) model, although we still keep the $k_x$-dependence of the wavefunctions. The approximation in Eq. 15 is important because it reduces the memory needed in the computation by 20X, allowing simulations of large NW cross sections with only minimal reduction in accuracy. Indeed, our numerical overlaps agree with the analytical expressions for the wavefunction overlaps if one assumes sine or cosine wavefunctions and parabolic bands, which can be derived to be $9/4A$ for intra-band and $1/A$ for inter-band transitions, where *A* is the cross section area of the NW [31, 40]. This is clearly indicated in Fig. 7, where we show the wavefunction overlap for the n-type [100] and [110] NWs with *D*=6nm between the state *k*=0, in subband *n*=1, and several final states in units of 1/*A*. In Fig. 7a, the intra-band transitions are shown with final states in subband *m*=1, and varying *k*-values. The wavefunction overlaps are indeed very close to the analytical value of 9/4. In Fig. 7b, the inter-band transitions are shown with final states in subbands $m=1,2,\ldots,12$ and $k=0$. The first point, for $m=1$ is the intraband transition which gives ~9/4, whereas for higher bands the overlaps reduce to lower values around ~1. The values are very close to the analytical ones, do not have significant $k$-dependence, and should not affect the qualitative nature of the results significantly. The price to pay, however, is that with this simplification the phase information for the wavefunctions is lost, and the selection rules are incorporated into the scattering rate calculation "by hand". However, this treatment is consistent with that for scattering in bulk and ultra-thin-layer structures reported in the literature. Still, even after this simplification, the storage of the probability density for the larger diameter NWs still requires several Giga bytes of memory.

For surface roughness (SR), we assume a 1D exponential autocorrelation function [50] for the roughness given by:

$$\langle \delta(\rho)\delta(\rho'-\rho)\rangle = \Delta_{rms}^2 e^{-\sqrt{2}|\rho|/L_C} \tag{16}$$

with $\Delta_{rms} = 0.48$nm and $L_C = 1.3$nm [48]. We derive the surface roughness matrix element assuming that SR only causes a band edge shift. The scattering strength is given by the shift in the subband edges with diameter scaling $\Delta E_{C,V}/\Delta D$ [51, 52]. The transition rate is derived as:

$$S_{n,m}^{SRS}\left(k_x,k_x^{'}\right) = \frac{2\pi}{\hbar}\left(\frac{q_0 \Delta E_{C,V}}{\Delta D}\right)^2 \left(\frac{2\sqrt{2}\Delta_{rms}^2 L_C}{2+q_x^2 L_C^2}\right)\delta\left(E_m\left(k_x^{'}\right) - E_n\left(k_x\right)\right), \tag{17}$$

where $q_x = k_x - k_x^{'}$. As described by various authors, the band edge variation is the cause of the major impact of SRS in ultra-scaled channels [48, 51, 52, 53, 54]. In Refs [48, 52] it was shown that that the SRS limited low-field mobility in ultra-thin nanostructures follows a $L^6$ behavior, where $L$ is the confinement length scale, originating from this subband shift due to the variation of $L$. This SRS model is a simplified one, compared to the ones described in Refs [48, 55, 56, 57] that account for additional Coulomb effects, the wavefunction deformation at the interface, and the position of electrons in the channel. These effects are ignored here since they only cause quantitative changes in our results, whereas our focus is on qualitative trends that originate from geometry-induced electronic structure variations.

Figure 8a and Fig. 8b show the shift in the band edges $\Delta E/\Delta D$ as a function of diameter for the conduction and valence subbands, respectively. Indeed, the trends follow a $D^{-3}$ power law both for electrons and holes as expected, with some minor deviations. For the n-type, the lowest valleys have slightly lower band edge shifts compared to the higher valleys. In the calculation of the SRS, the Γ and off-Γ valleys are taken separately into account when calculating the scattering rate. The orientation dependence is more evident in the case of p-type NWs. The band edge shifts are larger for the [100] NWs, whereas the band edges of the [111] NW are affected the least by diameter variations. The sensitivity of the band edges can be directly correlated with a confinement effective mass $m_C^*$. Using the simple notion of a particle in a box where the ground state energy is $E = \pi^2 \hbar^2 / 2 m_C^* D^2$, approximate values for $m_C^*$ can be extracted. These are shown in Fig. 8c and Fig. 8d. For n-type NWs, the [110] orientation shows the largest $m_C^*$, whereas for p-type NWs the [110] and [111] NWs have the largest $m_C^*$. The slight deviation in the band edges from the $D^{-3}$ law at smaller diameters, which reduces the rate of increase in the scattering matrix element, are also reflected as an increase in the confinement effective mass. The value of $m_C^*$ of the n-type NWs lies between the longitudinal and transverse bulk Si masses of $m_l=0.9m_0$ and $m_t=0.19m_0$. For p-type NWs, the $m_C^*$ values for the larger NW diameters are close to the bulk Si heavy-hole mass $m_{hh}=0.4m_0$. For the [100] orientation they remain in that region for the smaller diameters as well. For the [110] and [111] orientations on the other hand, $m_C^*$ increases as the diameter is reduced. This is an important observation that indicates that the p-type [111] and [110] NWs will be less sensitive to surface roughness scattering (SRS). For thermoelectric materials this can be especially important since SRS is needed for the reduction in thermal conductivity $\kappa_l$. The fact that an intrinsic bandstructure mechanism makes the conductivity more tolerant to SRS could help in power factor optimization in such channels in which rough boundaries are favored.

For ionized impurity scattering the scattering potential is approximated by:

$$U_S(\vec{r}) = \frac{q_0^2}{4\pi\kappa_s\varepsilon_0} \frac{e^{-\sqrt{(\vec{R}-\vec{R}')^2+x^2}/L_D}}{\sqrt{(\vec{R}-\vec{R}')^2+x^2}}, \quad (21)$$

where $\vec{R}$ is the position of an electron in the 2D cross section at $x=0$, influenced by an impurity at $(x,\vec{R}')$, and the $x$ direction is assumed to extend to infinity. The 3D screening length $L_D$ is given by:

$$L_D = \sqrt{\frac{\kappa_s\varepsilon_0 k_B T}{q_0^2 \mathrm{n}} \frac{\Im_{-1/2}(\eta_F)}{\Im_{-3/2}(\eta_F)}}. \quad (22)$$

where $\Im_\alpha(\eta_F)$ is the Fermi-Dirac integral of order $\alpha$, and $\mathrm{n}$ is the carrier concentration. The matrix element for electron-impurity scattering then becomes:

$$H^{m,n}_{k_x',k_x}(\vec{R}') = \int_R \frac{\langle F_{n,k_x}(\vec{R})\rangle}{\sqrt{A}} \left( \frac{1}{L_x} \int_{-\infty}^{\infty} \frac{q_0^2 e^{iq_x x}}{4\pi\kappa_s\varepsilon_0} \frac{e^{-\sqrt{(\vec{R}-\vec{R}')^2+x^2}/L_D}}{\sqrt{(\vec{R}-\vec{R}')^2+x^2}} dx \right) \frac{\langle F_{m,k_x'}(\vec{R})\rangle}{\sqrt{A}} d^2R \quad (23)$$

where the expression in the brackets is the Green's function of the infinitely long channel device. For a cylindrical channel, the expression in the parenthesis is the modified Bessel function of second kind of order zero, $K_0(q,\vec{R}')$ [48, 58, 59, 60].

The total transition rate due to impurity scattering is computed after taking the square of the matrix element, multiplying by $N_I L_x$, the number of impurities in the normalized cross sectional area of the NW in the length of the unit cell, and integrating over the distribution of impurities in the cross sectional area (over $\vec{R}'$). The impurities are assumed to be distributed uniformly in the volume considered.

The transport distribution function (TD) in Eq. 5 turns out to be a very convenient means to understand the effect of the electronic structure on the thermoelectric (TE) coefficients. Figure 9a shows the phonon-limited TDs for n-type NWs of $D$=3nm. The TDs for the three different orientations [100], [110] and [111] are shown. There are two observations that determine the performance of the NWs. i) The low energy linear region, where only one subband participates in transport, with slope proportional to $\eta/m^*$, where $\eta$ is the degeneracy of the subband [39]. ii) The separation of the TD from the Fermi level, $\eta_F$. The closer the TD is to the Fermi level for a particular carrier concentration and the higher its slope, the larger the conductivity and mobility will be. This is shown in Fig. 9b, where the orientation dependence of the mobility correlates with the order the TDs appear with respect to the Fermi level. It is important to note that since the transport and density-of-states (DOS) effective masses ($m^*$) are the same for NWs, a reduction in $m^*$ will not only reduce $\eta_F$ in order to keep the carrier concentration fixed, but it will also increase the TD slope, finally having a doubly positive impact on the conductivity [39, 61]. Figure 9c shows a different situation, regarding the TDs for p-type NWs of $D$=12nm in the three orientations. As the NW diameter increases, the DOS of the NWs approaches the bulk DOS, and $\eta_F$ is the same for all NWs. Their slope, however, is different, which is reflected in the large anisotropy in the mobility in Fig. 9d. Note that there are more subbands which result in more peaks in the TDs of the larger NWs compared to the narrower ones.

As mentioned previously, one of the approximations used is that of *bulk* phonons. Bulk phonons provide an ease of modeling and allow the understanding of the bandstructure effects on the TE coefficients, still with good qualitative accuracy in the results. Confined phonons in NWs can have very different dispersions and properties than bulk. However, the effect of phonon confinement for the thinnest NWs examined in this work is not that strong; it can be of the order of 10-20% (reduction in conductivity), and declines fast as the diameter increases [45, 46, 47, 62]. In the literature it is common to employ higher than bulk deformation potential values to account for phonon confinement [48, 63, 64, 65]. Here we use deformation potential parameters $D_{ODP}^{holes} = 13.24 \times 10^{10}$ eV/m, $D_{ADP}^{holes} = 5.34$ eV, and $D_{ADP}^{electrons} = 9.5$ eV from Refs [45, 46, 61] which are more suitable for NWs. All other electron-phonon coupling parameters are the bulk values taken from [40]. The qualitative behavior of our results mostly depends on the shape of the bandstructure and not on the strength of the phonon scattering mechanisms. For a more quantitative description of the results, phonon confinement has to be accounted for. However, even in that case, phonon scattering is not the major scattering mechanism in NW devices suited for TE applications. This is clearly demonstrated in Fig. 10, again using the TD features for ADP-limited (blue-dotted), ADP-ODP-limited (blue-solid), SRS-limited (red), impurity scattering-limited (green), and the TD including all scattering mechanisms (black). In Fig. 10a moderate values for surface roughness ($\Delta_{rms}$=0.24nm) and impurity concentration ($n_0=10^{18}$/cm$^3$) are used. Strong scattering will lower the TD value and degrade conductivity. From the important low energy region, we observe that both SRS and impurity scattering mechanisms are stronger than phonon scattering. Figure 10b shows the same features, but with $\Delta_{rms}$=0.48nm and $n_0=10^{19}$/cm$^3$, which are more relevant for high performance thermoelectric devices (around the peak of the power factor as it will be shown in section IV). SRS and impurity scattering are much stronger than phonon scattering. A calculation of the phonon contribution to the total scattering rate shows that it is only 12% and 6% in the situations of Fig. 10a and Fig. 10b, respectively, even with the larger than bulk deformation potential values [23]. The strongest mechanism is impurity scattering, which dominates the scattering processes at such high concentrations. Indeed, this is in agreement with impurity scattering in bulk Si which reduces the mobility by almost an order of magnitude from the phonon-limited value at such high concentrations [66]. This shows that the details of phonon scattering strength for NW devices might not be of great importance to the total channel conductivity. This also demonstrates the importance of modulation doping in achieving high thermoelectric performance.

## IV. Si nanowire thermoelectric coefficients

Geometrical features such as diameter and orientation will affect the electronic structure, and influence the electrical conductivity and the Seebeck coefficient. If one considers a specific carrier concentration, the influence of geometry shows up is two ways: i) The band edges (or transport distribution functions TD) shift with respect to the Fermi level as the geometry changes. ii) The effective masses (or carrier velocities) change. The changes will be different for different NW cases. A change in $\eta_F = E_F - E_C$ will affect both the conductivity and the Seebeck coefficient. This effect is shown in Fig. 11a and Fig. 11b, respectively, using a simple 1D subband and effective mass approximation. Changes in $\eta_F$ affect the conductivity exponentially, but affect the Seebeck coefficient only linearly, (and in an inverse way). The conductivity, therefore, is affected much more than the Seebeck coefficient. At a specific carrier concentration, changes in $\eta_F$ can happen as follows: i) In a NW channel with only a few subbands, once the diameter is reduced, $\eta_F$ increases in order to keep the carrier concentration constant as explained in detail in Refs [23, 39]. This reduces the conductivity exponentially. ii) The DOS changes through electronic structure modifications and $\eta_F$ will adjust to keep the carrier concentration constant.

As a consequence, since the electronic structures of the NWs in different orientations are different, $\eta_F$ will differ as well, resulting in orientation and geometry dependence of TE performance. Figure 12 shows the power factor for the n-type (solid) and p-type (dashed) NWs with $D$=10nm, in the [100] (blue), [110] (red), and [111] (green) transport orientations. The Boltzmann transport formalism was used. Some orientation dependence can be observed. Especially in p-type NWs the [111] orientation gives almost ~2X higher power factor than the other two p-type NW orientations. Note that p-type NWs perform lower than the n-type NWs for this NW diameter, but this difference is less severe for smaller diameters [23].

The conductivity usually degrades with diameter reduction because of the enhancement of scattering mechanisms such as phonon and surface roughness scattering (SRS) at smaller feature sizes. Figure 13 shows the effect of the diameter reduction on the TE coefficients for the [100] n-type NW at room temperature. Phonon scattering and SRS are considered. Figure 13a shows that the electrical conductivity decreases as the diameter of the NW is reduced. On the other hand, the Seebeck coefficient in Fig. 13b increases for the smaller diameters due to an $\eta_F$ increase. Overall, the power factor in Fig. 13c decreases with diameter, because the conductivity is degraded much more than the Seebeck coefficient is improved. Using an experimentally measured value for the thermal conductivity $\kappa_l$=2W/mK [15, 18], we compute the *ZT* figure of merit in Fig. 13d. *ZT* is reduced with diameter reduction, following the trend in the power factor. Two important observations can be made at this point: i) The conclusions are different from what previously described in Fig. 3 and Fig. 4 for ballistic transport. The increase in the power factor and *ZT* at reduced feature sizes is not observed when scattering is incorporated. On the contrary, the performance is degraded, because of a reduction in the conductivity. ii) *ZT* ~0.5-1 can be achieved in Si NWs, in agreement with recent experimental measurements [12, 13] (reduced from *ZT*~4 under ballistic considerations in Fig. 4). On the other hand, the value $\kappa_l$=2W/mK used for the calculation of *ZT* is measured for Si NWs of diameters *D*=15nm [15, 18]. This might be even smaller for smaller NW diameters or even orientation dependent [67, 68]. The power factor and *ZT* could potentially change and higher performance could be achieved. Nevertheless, the magnitude of these results is in agreement with other reports, both theoretical [37, 69] and experimental [12, 13, 14, 65].

The results in Fig. 13 only consider phonon scattering and SRS. The peak of the power factor, however, appears at carrier concentrations of $10^{19}$/cm$^3$. In order to reach such concentration high doping levels are required and the effect of impurity scattering thus cannot be excluded. In Fig. 14, we demonstrate the effect of different scattering mechanisms for the n-type [100] NW of diameters *D*=5nm. The conductivity in Fig. 14a is strongly degraded from the phonon-limited values (blue) once surface roughness scattering-SRS (black) and most importantly impurity scattering (red) are included in the calculation. The impurity concentration used at each instance is equal to the carrier concentration. The Seebeck coefficient in Fig. 14b does not change significantly with the introduction of additional scattering mechanisms because it is independent of scattering at first order [31]. The conductivity dominates the power factor, which is drastically reduced due to SRS and mostly impurity scattering (Fig. 14c). This can also reduce the *ZT* as shown in Fig. 14d from *ZT*~1 down to *ZT*~0.2. Since impurity scattering is such a strong mechanism, for high performance NW TEs alternative doping schemes need to be employed such as modulation doping or charge transfer techniques [65, 70, 71, 72].

## V. Conclusions

We presented a methodology that couples the atomistic $sp^3d^5s^*$-SO tight-binding model to two different transport formalisms: i) Landauer ballistic and ii) Linearized Boltzmann theory for calculating the thermoelectric power factor in ultra-thin Si nanowires. We introduced some approximations needed to make such methodology robust and efficient, and explained the differences in the conclusions obtained from these two different transport methods. Using this formalism the computational domain can be extended to "large" feature sizes (>10nm) still accounting for all atomistic effects, so that the length scale degree of freedom can be properly used as a design parameter. We show that geometrical features such as cross section and orientations could potentially provide optimization directions for the thermoelectric power factor in NWs. In the Si NWs investigated, low-dimensionality and geometrical features affect the electrical conductivity much more than the Seebeck coefficient. The conductivity is, therefore, the quantity that controls the behavior of the power factor and the figure of merit *ZT*, in contrast to the current view that the low-dimensional features could provide benefits through improvements in the Seebeck coefficient. We finally show that impurity scattering is the strongest scattering mechanism in nanowire thermoelectric channels, and ways that allow high carrier concentration without direct doping could largely improve the performance.

## Acknowledgements

This work was supported by the Austrian Climate and Energy Fund, contract No. 825467.

# APPENDIX 1: Wavefunction overlap integral / numerical calculation of the sum

The wavefunctions in tight-binding are sampled on the atomic sites. Equations (11) and (23) involve integrations over the in-plane $R$ perpendicular to the nanowire axis. In the calculation the integrals over $R$ are performed by transforming the integrals to summations over the atomic sites $N$. Below we demonstrate how this is performed for the calculation of the wavefunction overlaps in the case of phonon scattering. The matrix element needs to be squared in the calculation of the scattering rates. What is required is integration of the type:

$$\frac{1}{A^{m,n}_{k_x',k_x}} = \frac{1}{A^2} \int_R \rho^{m,n}_{k_x',k_x}(\vec{R})^2 \, d^2R, \tag{A1.1a}$$

where
$$\rho^{m,n}_{k_x',k_x}(\vec{R}) = F_{m,k_x'}(\vec{R})^* F_{n,k_x}(\vec{R}) \tag{A1.1b}$$

and $1/A^2$ originates from wavefunction normalization.

We convert the integral to a sum by

$$\frac{1}{A} \int_R (\cdot) \, d^2R \approx \frac{1}{A} \sum_R (\cdot) \Delta A = \frac{1}{N} \sum_R (\cdot), \tag{A1.2a}$$

where $\Delta A = A/N$, and $N$ is the number of atomic sites in the unit cell of the NW. Therefore,

$$\frac{1}{A^{m,n}_{k_x',k_x}} = \frac{1}{A^2} \int_R \rho^{m,n}_{k_x',k_x}(\vec{R})^2 \, d^2R = \frac{1}{AN} \sum_R \rho^{m,n}_{k_x',k_x}(\vec{R})^2 = \frac{1}{A}\left(\frac{1}{N}\sum_R \rho^{m,n}_{k_x',k_x}(\vec{R})^2\right). \tag{A1.3}$$

In the wavefunction normalization, the usual expression in integral or summation form is:

$$\frac{1}{A} \int_R F_{n,k_x}(\vec{R}) F_{n,k_x}(\vec{R})^* \, dA = 1, \tag{A1.4a}$$

or
$$\frac{1}{N} \sum_R F_{n,k_x}(\vec{R}) F_{n,k_x}(\vec{R})^* = 1. \tag{A1.4b}$$

Numerically, however, the wavefunctions provided by the eigenvalue solvers are already normalized and give:

$$\sum_R \tilde{F}_{n,k_x}(\vec{R}) \tilde{F}_{n,k_x}(\vec{R})^* = 1. \tag{A1.5}$$

where $F_{n,k_x} = \sqrt{N} \tilde{F}_{n,k_x}$.

The expression in Eq. (A1.3) then becomes:

$$\frac{1}{A^{m,n}_{k_x',k_x}} = \frac{1}{A}\left(N \sum_R \tilde{\rho}^{m,n}_{k_x',k_x}(\vec{R})^2\right),$$

where $\tilde{\rho}^{m,n}_{k_x',k_x}(\vec{R})$ is calculated using the actual $F$ expressions given by the eigenvalue solver, and: $\rho^{m,n}_{k_x',k_x}(\vec{R}) = N^2 \tilde{\rho}^{m,n}_{k_x',k_x}(\vec{R})$.

## APPENDIX 2: Bond Passivation (sp$^3$ - Hybridization) in tight-binding, following for Ref. [36]:

The passivation of the bonds that reside outside the domain of the NW, is done using a $sp^3$ hybridized scheme. The construction of the Hamiltonian assumes sums of couplings between atomic basis orbitals (orbital-space). This means that each on-site element represents a specific orbital and has contributions from four bonds (couplings). In order to passivate a specific bond, a transformation to the hybridized $sp^3$ space is performed. This means that the transformed matrix reside in the hybridized "bond-space", in which all hybridized orbitals are aligned along the four bond directions. The on-side element of the hybridized orbital along the dangling bond direction that is to be passivated is then raised to a large value (30 eV), in order to be placed away from the energies of interest and not to affect the bandstructure calculation. The bonds from an anion to the four cations and vice versa, are formed primarily by $sp^3$-hybridization as a linear combination of only the $s$ and $p$ orbitals. The $sp^3$ hybridized orbitals from an anion to the cations are:

$$\begin{aligned}
|sp^3\rangle_{[111]}^{a \to c} &= \frac{1}{2}(|s\rangle + |p_x\rangle + |p_y\rangle + |p_z\rangle) \\
|sp^3\rangle_{[\bar{1}\bar{1}1]}^{a \to c} &= \frac{1}{2}(|s\rangle - |p_x\rangle - |p_y\rangle + |p_z\rangle) \\
|sp^3\rangle_{[1\bar{1}\bar{1}]}^{a \to c} &= \frac{1}{2}(|s\rangle + |p_x\rangle - |p_y\rangle - |p_z\rangle) \\
|sp^3\rangle_{[\bar{1}1\bar{1}]}^{a \to c} &= \frac{1}{2}(|s\rangle - |p_x\rangle + |p_y\rangle - |p_z\rangle)
\end{aligned}
\Rightarrow
\begin{bmatrix}
|sp^3\rangle_{[111]}^{a \to c} \\
|sp^3\rangle_{[\bar{1}\bar{1}1]}^{a \to c} \\
|sp^3\rangle_{[1\bar{1}\bar{1}]}^{a \to c} \\
|sp^3\rangle_{[\bar{1}1\bar{1}]}^{a \to c}
\end{bmatrix}
= \frac{1}{2}
\begin{bmatrix}
1 & 1 & 1 & 1 \\
1 & -1 & -1 & 1 \\
1 & 1 & -1 & -1 \\
1 & -1 & 1 & -1
\end{bmatrix}
\begin{bmatrix}
|s\rangle \\
|p_x\rangle \\
|p_y\rangle \\
|p_z\rangle
\end{bmatrix}
= V_{sp^3}^{a \to c}$$

(A2.1)

whereas the $sp^3$ hybridized orbitals from a cation to the four anions are:

$$\begin{aligned}
|sp^3\rangle_{[\bar{1}\bar{1}\bar{1}]}^{c \to a} &= \frac{1}{2}(|s\rangle - |p_x\rangle - |p_y\rangle - |p_z\rangle) \\
|sp^3\rangle_{[11\bar{1}]}^{c \to a} &= \frac{1}{2}(|s\rangle + |p_x\rangle + |p_y\rangle - |p_z\rangle) \\
|sp^3\rangle_{[\bar{1}11]}^{c \to a} &= \frac{1}{2}(|s\rangle - |p_x\rangle + |p_y\rangle + |p_z\rangle) \\
|sp^3\rangle_{[1\bar{1}1]}^{c \to a} &= \frac{1}{2}(|s\rangle + |p_x\rangle - |p_y\rangle + |p_z\rangle)
\end{aligned}
\Rightarrow
\begin{bmatrix}
|sp^3\rangle_{[\bar{1}\bar{1}\bar{1}]}^{c \to a} \\
|sp^3\rangle_{[11\bar{1}]}^{c \to a} \\
|sp^3\rangle_{[\bar{1}11]}^{c \to a} \\
|sp^3\rangle_{[1\bar{1}1]}^{c \to a}
\end{bmatrix}
= \frac{1}{2}
\begin{bmatrix}
1 & -1 & -1 & -1 \\
1 & 1 & 1 & -1 \\
1 & -1 & 1 & 1 \\
1 & 1 & -1 & 1
\end{bmatrix}
\begin{bmatrix}
|s\rangle \\
|p_x\rangle \\
|p_y\rangle \\
|p_z\rangle
\end{bmatrix}
= V_{sp^3}^{c \to a}$$

(A2.2)

The passivation is then achieved by a transformation as follows:

$$[H]_{Hybrid} = V_{sp^3} [H_{E(sp)}] V_{sp^3}^{\dagger},$$

(A2.3)

where: $H_{E(sp)} = \begin{bmatrix} E_s & & & \\ & E_{px} & & \\ & & E_{py} & \\ & & & E_{pz} \end{bmatrix}$, is the on-site matrix consisting only of the *s* and

*p* orbitals. Once the transformation takes place, the on-site elements of the hybridized space matrix along the bonds to be passivated are raised by $(h_{sp3})_{i,i}=30eV$. Finally, a back transformation into the orbital space will give the passivated matrix elements:

$$[H]_{Passiv.} = V_{sp^3}^\dagger \left[ H_{Hybrid} + h_{sp^3} \right] V_{sp^3} \text{, where} \quad (A2.4)$$

$h_{sp^3} = \begin{bmatrix} a_1 & & & \\ & a_2 & & \\ & & a_3 & \\ & & & a_4 \end{bmatrix}$, with $a_i$ been 30eV or zero, depending on whether the bond *i* is passivated or not.

# References

[1]  G. J. Snyder and E. S. Toberer, Nature Materials, vol. 7, pp. 105–114, 2008.

[2]  A. Majumdar, Science Materials, vol. 303, pp. 777–778, 2004.

[3] L.D. Hicks, and M. S. Dresselhaus, Phys. Rev. B, vol. 47, no. 24, p. 16631, 1993.

[4] R. Venkatasubramanian, E. Siivola, T. Colpitts, and B. O' Quinn, Nature, vol. 413, pp. 597-602, 2001.

[5]  D. A. Broido and T. L. Reinecke, Phys. Rev. B, vol. 51, p. 13797, 1994.

[6]  T. C. Harman, P. J. Taylor, M. P. Walsh, and B. E. LaForge, Science, 297, 2229–2232, 2002.

[7] M. Dresselhaus, G. Chen, M. Y. Tang, R. Yang, H. Lee, D. Wang, Z. Ren, J.-P. Fleurial, and P. Gagna, Adv. Mater., vol. 19, pp. 1043-1053, 2007.

[8] A. Shakouri, IEEE International Conference on thermoelectrics, pp. 495–500, 2005.

[9] J. O. Sofo and G. D. Mahan, Appl. Phys. Lett., vol. 65, p. 2690 (3pp), 1994.

[10] W. Kim, S. L. Singer, A. Majumdar, D. Vashaee, Z. Bian, A. Shakouri, G. Zeng, J. E. Bowers, J. M. O. Zide, and A. C. Gossard, Appl. Phys. Lett., vol. 88, p. 242107, 2006.

[11] Y. Wu, R. Fan, and P. Yang, Nano Lett., vol. 2, no. 2, pp. 83-83, 2002.

[12] A. I. Boukai, Y. Bunimovich, J. T.-Kheli, J.-K. Yu, W. A. G. III, and J. R. Heath, Nature, 451, 168–171, 2008.

[13] A. I. Hochbaum, R. Chen, R. D. Delgado, W. Liang, E. C. Garnett, M. Najarian, A. Majumdar, and P. Yang, Nature, vol. 451, pp. 163–168, 2008.

[14] J. Tang, H.-T. Wang, D. H. Lee, M. Fardy, Z. Huo, T. P. Russell, and P. Yang, Nano Lett., 10, 10, 4279-4283, 2010.

[15] D. Li, Y. Wu, R. Fang, P. Yang, and A. Majumdar Appl. Phys. Lett., 83, 3186–3188, 2003.

[16] K. Nielsch, J. Bachmann , J. Kimling , and H. Böttner, Adv. Energy Mater., 1, 713, 2011.

[17] G. Chen, Semiconductors and Semimetals, vol. 71, pp. 203–259, 2001.

[18]  R. Chen, A. I. Hochbaum, P. Murphy, J. Moore, P. Yang, and A. Majumdar, Phys. Rev. Lett., vol. 101, p. 105501, 2008.

[19] D. Li, S. T. Huxtable, A. R. Abramsin, and A. Majumdar, Trans. of the ASME, 127, 108–114, 2005.

[20] P. Martin, Z. Aksamija, E. Pop, and U. Ravaioli, Phys. Rev. Lett., 102, 125503, 2009.

[21] C. J. Vineis, A. Shakouri, A. Majumdar, and M. C. Kanatzidis, Adv. Mater., 22, 3970-3980, 2010.

[22] C. M. Jaworski, V. Kulbachinskii, and J. P. Heremans, Phys. Rev. B, 80, 125208, 2009.

[23] N. Neophytou and H. Kosina, Phys. Rev. B, vol. 83, 245305, 2011.

[24]  G. D. Mahan and J. O. Sofo, Proc. Natl. Acad. Sci. USA, vol. 93, pp. 7436-7439, 1996.

[25]   T. B. Boykin, G. Klimeck, and F. Oyafuso, Phys. Rev. B, vol. 69, no. 11, pp. 115201-115210, 2004.

[26]  G. Klimeck, S. Ahmed, B. Hansang, N. Kharche, S. Clark, B. Haley, S. Lee, M. Naumov, H. Ryu, F. Saied, M. Prada, M. Korkusinski, T. B. Boykin, and R. Rahman, IEEE Trans. Electr. Dev., vol. 54, no. 9, pp. 2079-2089, 2007.

[27]  G. Klimeck, S. Ahmed, N. Kharche, M. Korkusinski, M. Usman, M. Prada, and T. B. Boykin, IEEE Trans. Electr. Dev., vol. 54, no. 9, pp. 2090-2099, 2007.

[28]  N. Neophytou, A. Paul, M. Lundstrom, and G. Klimeck, IEEE Trans. Elect. Dev., vol. 55, no. 6, pp. 1286-1297, 2008.

[29] N. Neophytou, A. Paul, and G. Klimeck, IEEE Trans. Nanotechnol., vol. 7,  no. 6, pp. 710-719, 2008.

[30] R. Landauer, IBM J. Res. Dev., vol. 1, p. 223, 1957.

[31]   R. Kim, S. Datta, and M. S. Lundstrom, J. Appl. Phys., vol. 105, p. 034506, 2009.

[32] C. Jeong, R. Kim, M. Luisier, S. Datta, and M. Lundstrom, J. Appl. Phys. 107, 023707 2010.

[33] G. Liang, W. Huang, C. S. Koong, J.-S. Wang, and J. Lan, J. Appl. Phys. 107, 014317, 2010.

[34]  N. Neophytou, M. Wagner, H. Kosina, and S. Selberherr, J. Electr. Materials, vol. 39, no. 9, pp. 1902-1908, 2010.

[35] T. J. Scheidemantel, C. A.-Draxl, T. Thonhauser, J. V. Badding, and J. O. Sofo, Phys. Rev. B, vol. 68, p. 125210, 2003.

[36] S. Lee, F. Oyafuso, P. Von, Allmen, and G. Klimeck, Phys. Rev. B, vol. 69, pp. 045316-045323, 2004.

[37]  T. T.M. Vo, A. J. Williamson, V. Lordi, and G. Galli, Nano Lett., vol. 8, no. 4, pp. 1111-1114, 2008.

[38] T. Markussen, A.-P. Jauho, and M. Brandbyge, Phys. Rev. Lett., vol. 103, p. 055502, 2009.

[39] N. Neophytou and H. Kosina, Phys. Rev. B, vol. 84, p. 085313, 2011.

[40] M. Lundstrom, "Fundamentals of Carrier Transport," Cambridge University Press, 2000.

[41] D. K. Ferry and S. M. Goodnick, "Transport in Nanostructures," Cambridge University Press, 1997.

[42] M. V. Fischetti, J. Appl. Phys., 89, 1232, 2001.

[43] M. V. Fischetti, Z. Ren, P. M. Solomon, M. Yang, and K. Rim, J. Appl. Phys., 94, 1079, 2003.

[44] D. Esseni and A. Abramo, IEEE Trans. Electr. Dev., 50, 7, 1665, 2003.

[45] A. K. Buin, A. Verma, A. Svizhenko, and M. P. Anantram, Nano Lett., vol. 8, no. 2, pp. 760-765, 2008.

[46]  A. K. Buin, A. Verma, and M. P. Anantram, J. Appl. Phys., vol. 104, p. 053716, 2008.

[47] E. B. Ramayya, D. Vasileska, S. M. Goodnick, and I. Knezevic, J. Appl. Phys., vol. 104, p. 063711, 2008.

[48] S. Jin, M. V. Fischetti, and T. Tang, Jour. Appl. Phys., 102, 83715, 2007.

[49] C. Jacoboni and L. Reggiani, Rev. Mod. Phys., vol. 55, 645, 1983.

[50] S. M. Goodnick, D. K. Ferry, C. W. Wilmsen, Z. Liliental, D. Fathy, and O. L. Krivanek, Phys. Rev. B, vol. 32, p. 8171, 1985.

[51] H. Sakaki, T. Noda, K. Hirakawa, M. Tanaka, and T. Matsusue, Appl. Phys. Lett., vol 51, no. 23, p. 1934, 1987.

[52] K. Uchida and S. Takagi, Appl. Phys. Lett., vol. 82, no. 17, pp. 2916-2918, 2003.

[53] T. Fang, A. Konar, H. Xing, and D. Jena, Phys. Rev. B, 78, 205403, 2008.

[54] J. Wang, E. Polizzi, A. Ghosh, S. Datta, and M. Lundstrom, Appl. Phys. Lett., 87, 043101, 2005.

[55] R. E. Prange and T.-W. Nee, Phys. Rev., vol. 168, 779, 1968.

[56] T. Ando, A. Fowler, and F. Stern, Rev. Mod. Phys., vol. 54, p. 437, 1982.

[57] D. Esseni, IEEE Trans. Electr. Dev., vol. 51, no. 3, p. 394, 2004.

[58] J. Lee and H. N. Spector, J. Appl. Phys., vol. 57, no. 2, p. 366, 1985.

[59] S. D. Sarma and W. Lai, Phys. Rev. B, vol. 32, no. 2, p. 1401, 1985.

[60] P. Yuh and K. L. Wang, Appl. Phys. Lett., vol. 49, no. 25, p. 1738, 1986.

[61] N. Neophytou and H. Kosina, Nano Lett., vol. 10, no. 12, pp. 4913-4919, 2010.

[62] L. Donetti, F. Gamiz, J. B. Roldan, and A. Godoy, J. Appl. Phys., vol. 100, p. 013701, 2006.

[63] M. V. Fischetti and S. E. Laux, J. Appl. Phys., 80, 2234, 1996.

[64] T. Yamada and D. K. Ferry, Solid-State Electron., 38, 881, 1995.

[65] H. J. Ryu, Z. Aksamija, D. M. Paskiewicz, S. A. Scott, M. G. Lagally, I. Knezevic, and M. A. Eriksson, Phys. Rev. Lett., vol. 105, p. 256601, 2010.

[66] H. Kosina and G. Kaiblinger-Grujin, Solid-State Electronics, vol. 42, no. 3, pp. 331-338, 1998.

[67] T. Markussen, A.-P. Jauho, and M. Brandbyge, Nano Lett., vol. 8, no. 11, pp. 3771-3775, 2008.

[68] Z. Aksamija and I. Knezevic, Phys. Rev. B, 82, 045319, 2010.

[69] T. Markussen, A.-P. Jauho, and M. Brandbyge, Phys. Rev. B, 79, 035415, 2009.

[70] S. Goswami, C. Siegert, S. Shamim, M. Pepper, I. Farrer, D. A. Ritchie, and A. Ghosh, Appl. Phys. Lett., 97, 132104, 2010.

[71] P. Zhang, E. Tewaarwerk, B.-N. Park, D. E. Savage, G. K. Celler, I. Knezevic, P. G. Evans, M. A. Eriksson, and M. G. Lagally, Nature Lett., 439, 703, 2006.

[72] T. He, J. He, M. Lu, B. Chen, H. Pan, W. F. Reus, W. M. Nolte, D. P. Nackashi, P. D. Franzon, and J. M. Tour, J. Am. Chem. Soc., 128, 14537, 2006.

Figure 1:

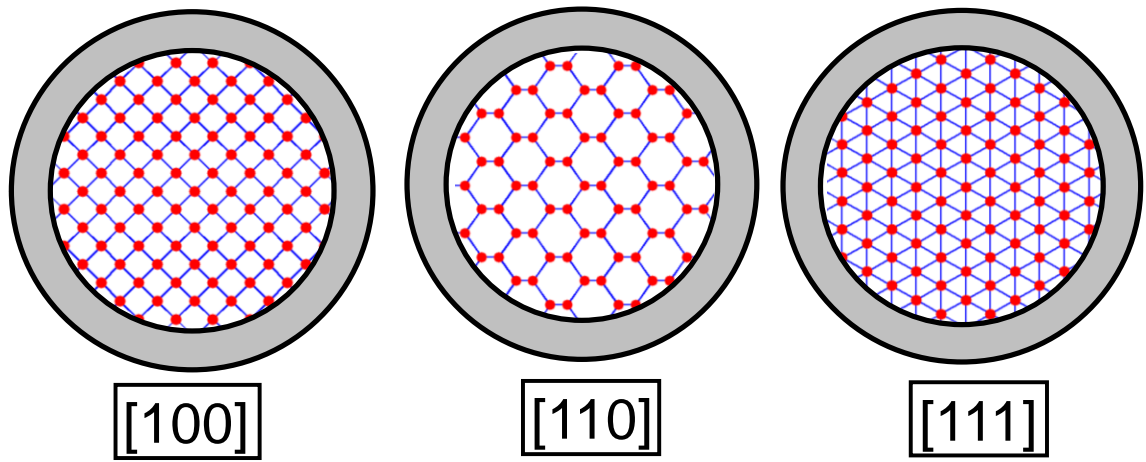

Figure 1 caption:

Zincblende lattice of cylindrical nanowires in the [100], [110], and [111] orientations.

Figure 2:

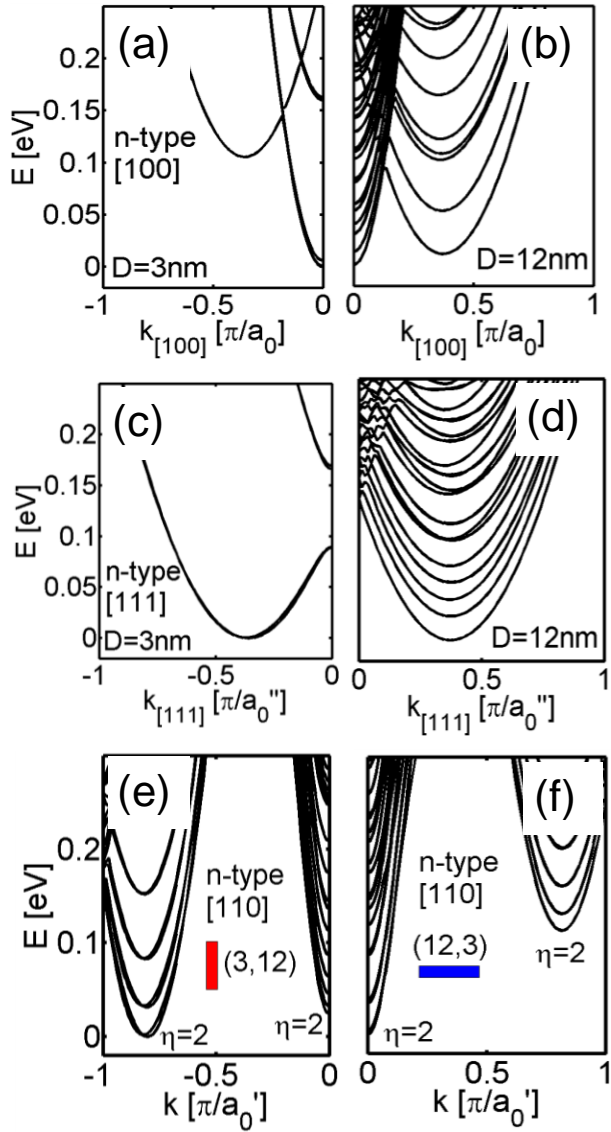

Figure 2 caption:

Dispersions of n-type NWs in various orientations and diameters/side lengths. (a) [100], $D$=3nm. (b) [100], $D$=12nm. (c) [111], $D$=3nm. (d) [111], $D$=12nm. (e-f) Dispersions of rectangular NWs with widths ($W$) and heights ($H$): (e) [110] NW, $W$=3nm, $H$=12nm. (f) [110] NW, $W$=12nm, $H$=3nm. $a_0$, $a_0'$ and $a_0''$ are the unit cell lengths for the wires in the [100], [110], and [111] orientations, respectively. The filled rectangles indicate the NW cross section.

Figure 3:

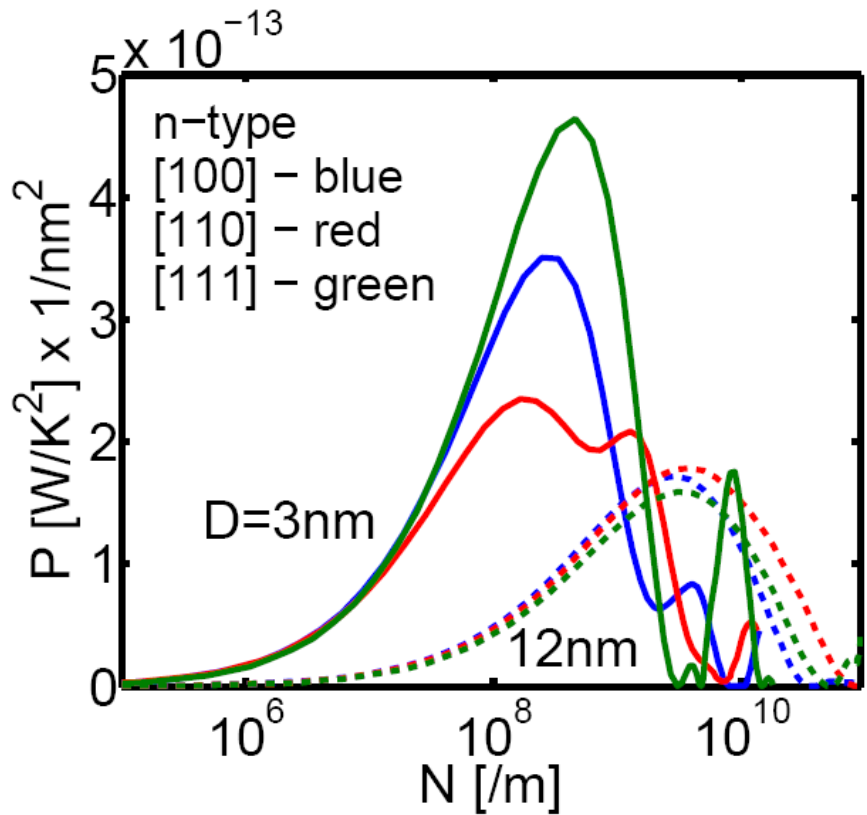

Figure 3 caption:

The thermoelectric power factor versus the 1D carrier concentration under ballistic transport conditions for n-type NWs of $D$=3nm and $D$=12nm in the [100] (blue), [110] (red), and [111] (green) orientations.

Figure 4:

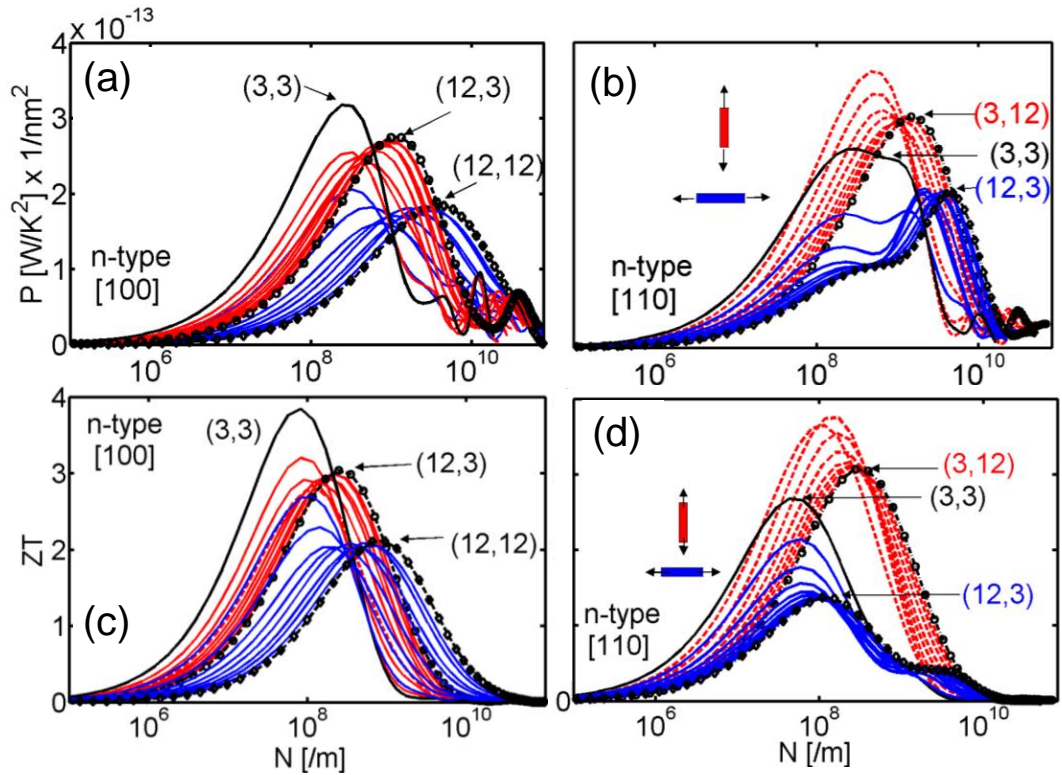

Figure 4 caption:

Thermoelectric features for n-type [100] (left column) and [110] (right column) NWs versus the 1D carrier concentration under ballistic transport conditions. (a, c) [100] NWs. Red lines: NWs with cross section sizes $W$=3nm to 12nm, while $H$=3nm fixed (wide and thin NWs, approaching a thin-body). Blue lines: square NWs with cross section sizes $W$=$H$=3nm to $W$=$H$=12nm. Increments in sizes are of 1nm. (a) Power factor $\sigma S^2$. (c) ZT figure of merit. (b, d) [110] NWs. Red lines: NWs with cross section sizes $W$=3nm fixed, and $H$=3nm to 12nm (thin and tall NWs, approaching a thin-body device). Blue lines: NWs with cross section sizes $W$=3nm to 12nm and $H$=3nm fixed (thin and wide NWs, approaching a thin-body). Increments in sizes are of 1nm. (b) Power factor $\sigma S^2$. (d) The ZT figure of merit. The filled rectangles indicate the NW cross section. $\kappa_l$=2W/mK is used for the thermal conductivity.

Figure 5:

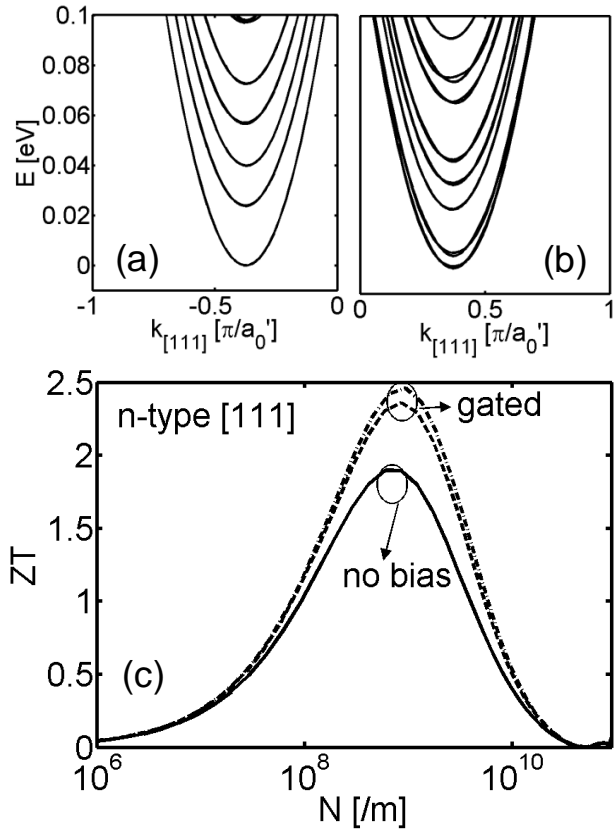

Figure 5 caption:

The effect of gate field electrostatic potential on the electronic structure and the ballistic *ZT* figure of merit of the n-type [111] D=12nm NW. (a) The bandstructure for a flat potential profile. (b) The bandstructure under strong inversion, for $V_G$-$V_T$=1.0V. (c) The *ZT* figure of merit for the NW versus carrier concentration under no bias, and under large gate bias cases.

Figure 6:

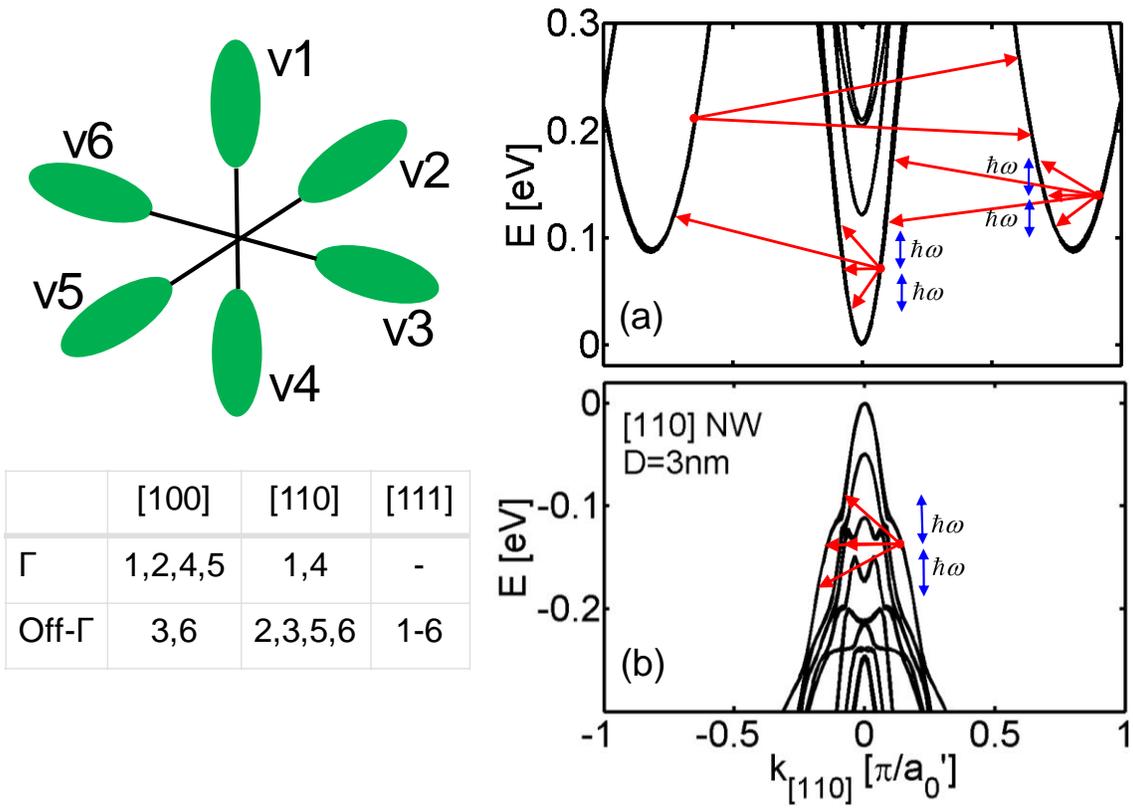

Figure 6 caption:

Dispersions of the [110] NW of $D$=3nm with the scattering mechanisms indicated. (a) n-type NW. Intra-valley elastic and inter-valley inelastic (IVS) processes are considered (between the three valleys), following the bulk silicon scattering selection rules. For NWs in different orientations the $\Gamma$ and off-$\Gamma$ valley degeneracies, and the bulk valleys from where they originate are shown in the table. Following the bulk scattering selection rules, however, each of the valleys is considered independently. (b) p-type NW. Elastic and inelastic processes are considered within the entire bandstructure. Intra- and inter-valley scattering is considered.

Figure 7:

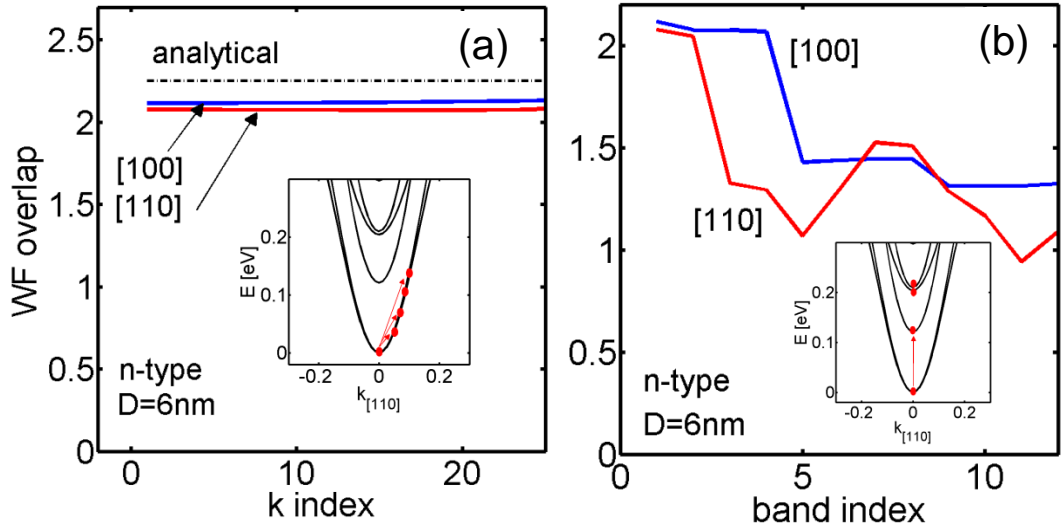

Figure 7 caption:

The wavefunction overlap integral between a state at k=0 in the first subband with (a) states of different *k* in the first subband, and (b) with states at *k*=0 but different subbands in units of 1/A, where A is the area of the NW. Results for n-type [100] and [110] NWs of *D*=6nm are shown. The analytical value for the integral is 9/4 for intra-band transitions, and 1 for inter-band transitions. Insets: Schematics indicating the initial and final *E(k)* states (the dispersions are of the *D*=3nm, [110] n-type NW, for which the transitions can be more easily visualized).

Figure 8:

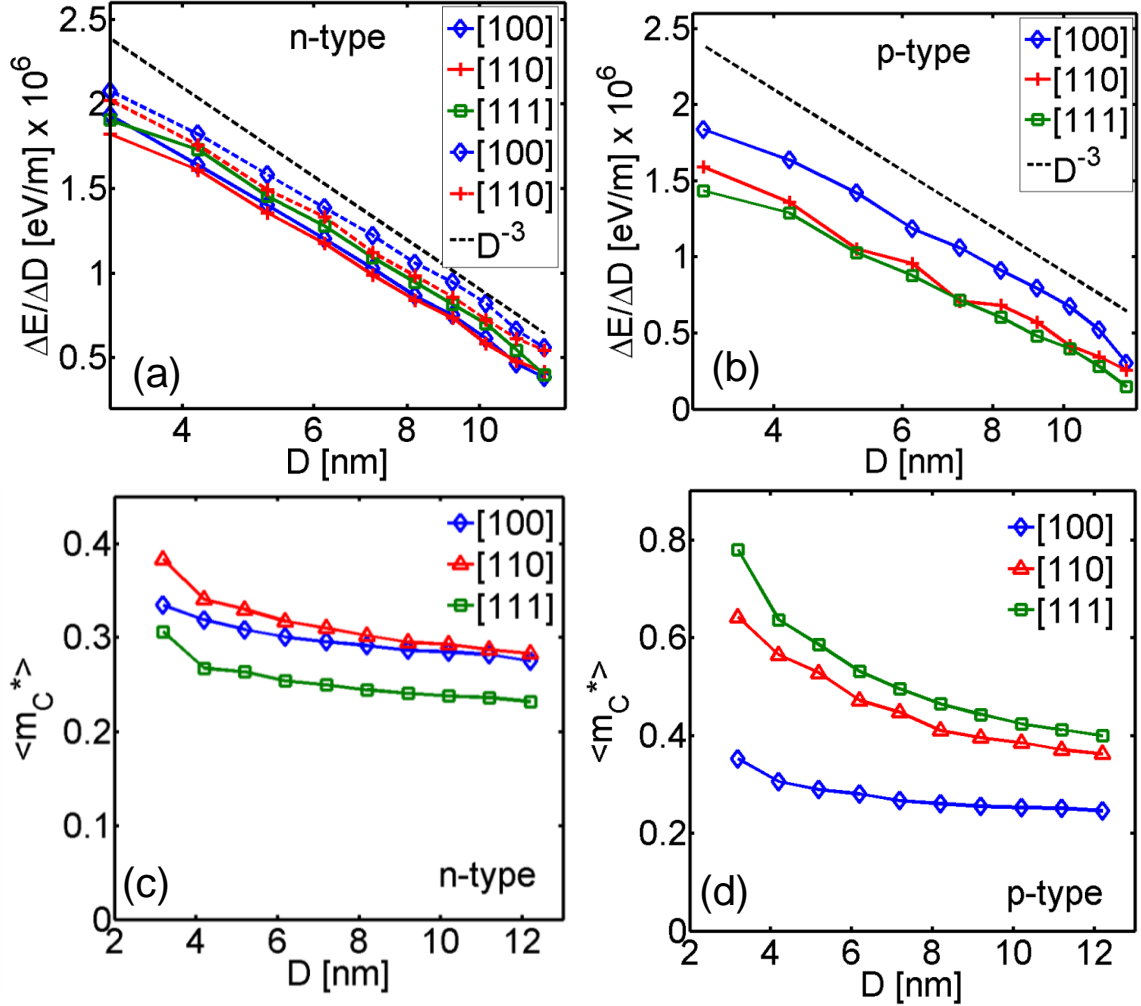

Figure 8 caption:

Change in band edges as a function of diameter. Results for NWs in the [100] (diamond-blue), [110] (cross-red), and [111] (square-green) transport orientations are shown. (a) Conduction band. Results for the lower valleys (solid), and upper valleys (dashed) are shown. (b) Valence band. The dashed-black line indicates the $D^{-3}$ law. (c-d) The average confinement effective mass for NWs in different orientations versus the diameter. This is calculated from the change in the subband edges with confinement using the particle in a box quantization picture. Results for [100] (diamond-blue), [110] (triangle-red) and [111] (square-green) transport orientated NWs are shown. (c) n-type NWs. (d) p-type NWs.

Figure 9:

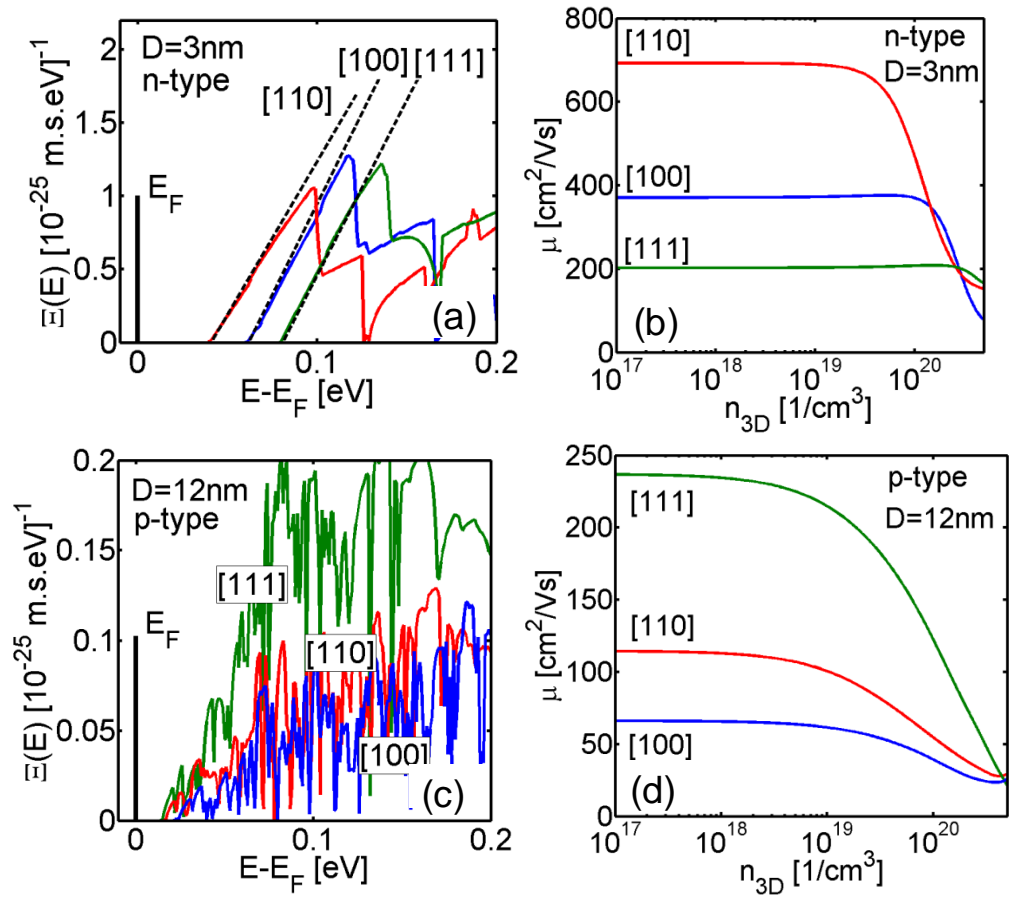

Figure 9 caption:

Low-field, phonon-limited transport characteristics for NWs in the [100] (blue), [110] (red) and [111] (green) orientations. (a) The transport distribution function (TD) of the $D$=3nm n-type NWs. (b) The mobility of $D$=3nm n-type NWs. (c) The transport distribution function (TD) of $D$=12nm p-type NWs. (d) The mobility of $D$=12nm p-type NWs.

Figure 10:

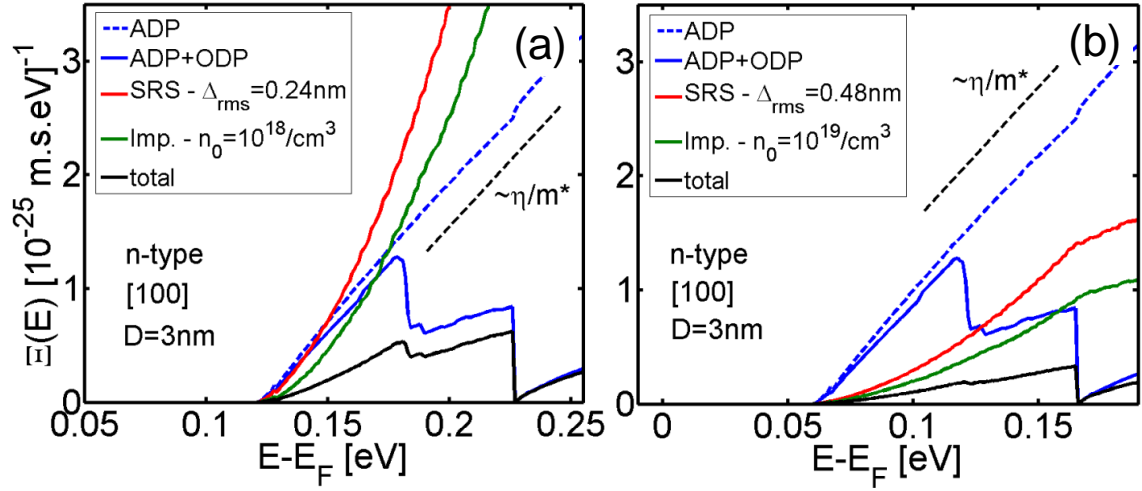

## Figure 10 caption:

The transport distribution function (TD) $\Xi(E)$ for n-type [100] NW of $D$=3nm under different scattering conditions i) ADP phonon-limited (blue dashed), ii) ADP-IVS phonon-limited (blue), iii) SRS limited (red), and iv) impurity limited (green). (a) Carrier concentration n=$10^{18}$/cm$^3$, $\Delta_{rms}$=0.24nm. (b) Carrier concentration n=$10^{19}$/cm$^3$, $\Delta_{rms}$=0.48nm. The impurity concentration is equal to the carrier concentration in each case.

Figure 11:

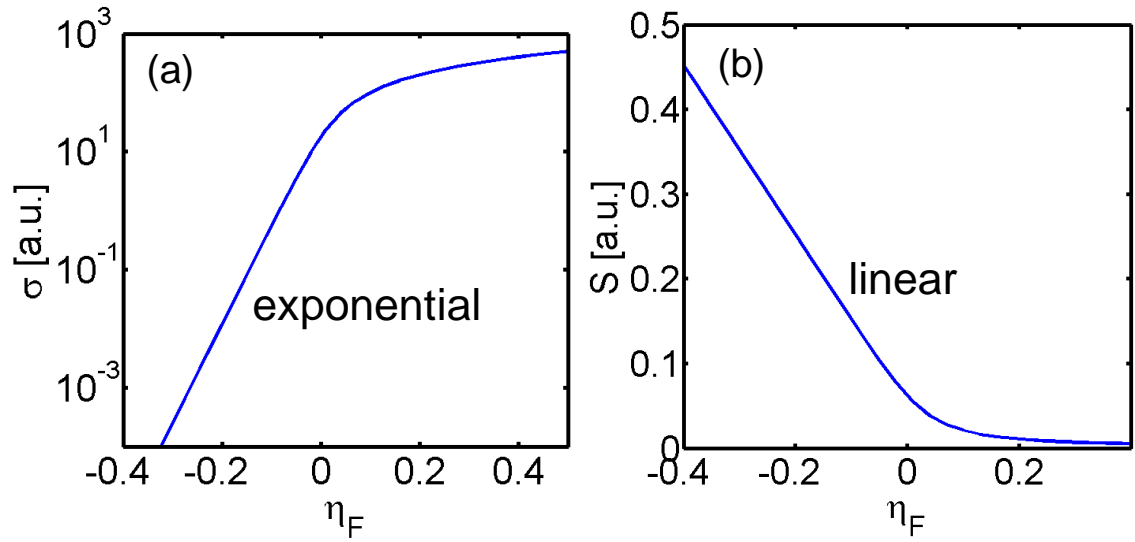

Figure 11 caption:

The electrical conductivity (a) and Seebeck coefficient (b) versus the distance of the conduction band from the Fermi level, $\eta_F = E_F - E_C$. A simple parabolic band and scattering rates proportional to the density of final states are assumed.

Figure 12:

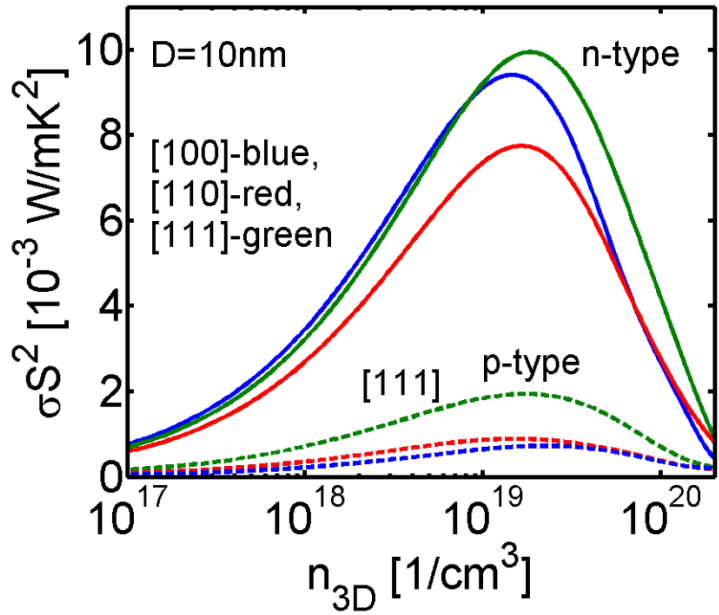

Figure 12 caption:

The phonon-limited thermoelectric power factor for $D=10$nm, n-type (solid) and p-type (dashed) NWs of different transport orientations versus the carrier concentration. Orientations are [100] (blue), [110] (red), and [111] (green).

Figure 13:

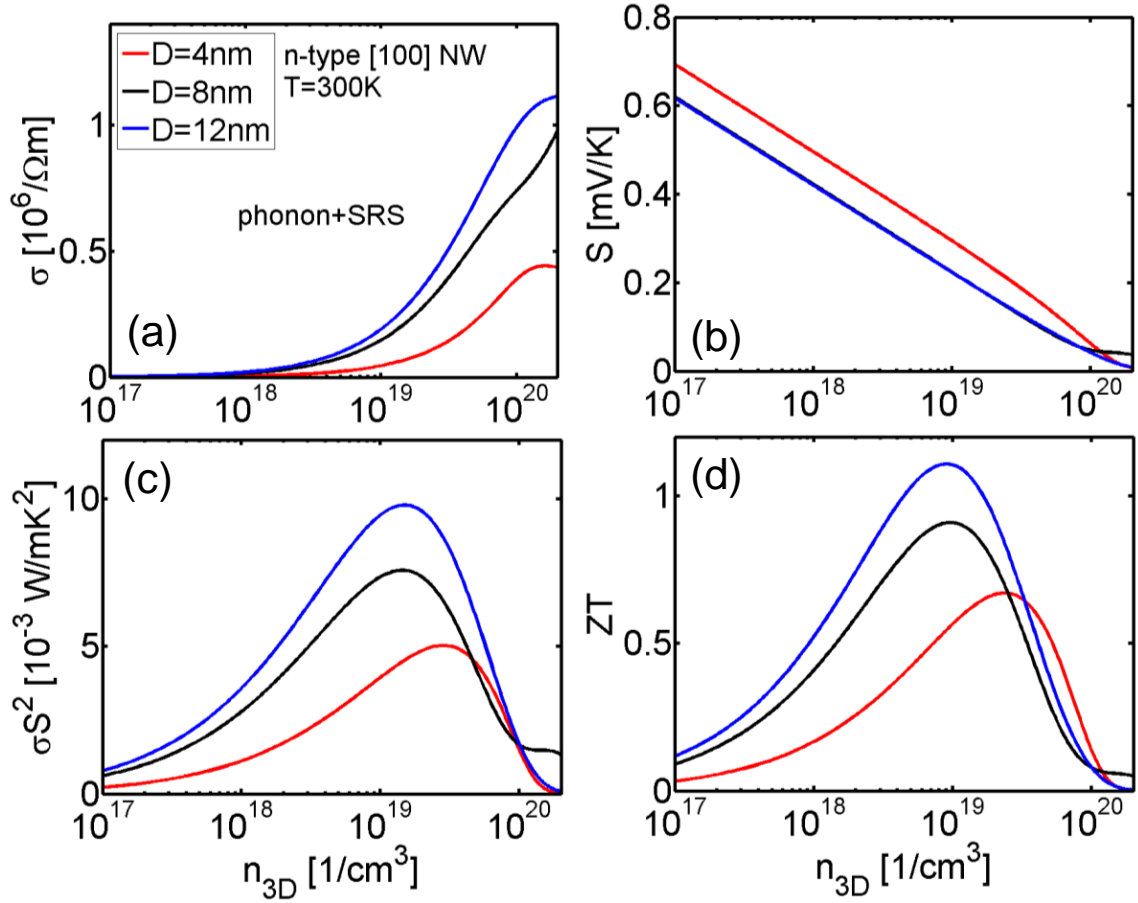

Figure 13 caption:

Thermoelectric coefficients versus carrier concentration for n-type [100] NWs of *D*=4nm (red), 8nm (black) and 12nm (blue), at 300K. Phonon scattering plus SRS are included. (a) The electrical conductivity. (b) The Seebeck coefficient. (c) The power factor. (d) The *ZT* figure of merit.

Figure 14:

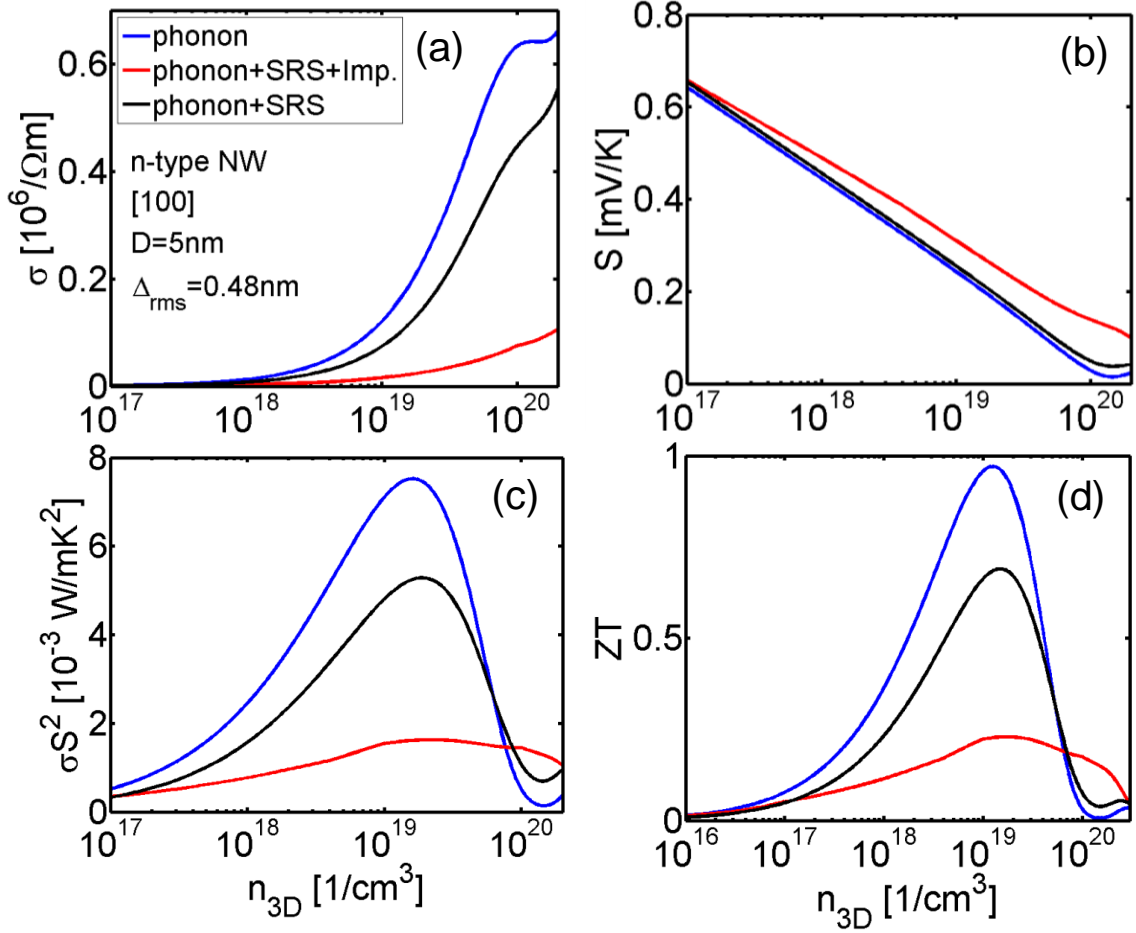

## Figure 14 caption:

Thermoelectric coefficients versus carrier concentration for a D=5nm, n-type NW, in the [100] transport orientation at 300K. Different scattering mechanisms are considered, phonons (blue), phonons plus SRS (black), and phonon plus SRS plus impurity scattering (red). For impurity scattering, $n_0 = n_{3D}$ is assumed. (a) The electrical conductivity. (b) The Seebeck coefficient. (c) The power factor. (d) The *ZT* figure of merit.